\documentclass{aa}
\usepackage{amsmath,amssymb}
\usepackage{txfonts}
\usepackage{latexsym}
\usepackage{graphicx}

\begin{document}

\title{Kerr Geodesics, the Penrose Process and Jet Collimation by a Black
Hole}
\author{ J. Gariel  \inst{1}\thanks{jerome.gariel@upmc.fr}
         \and  M.A.H. MacCallum \inst{2}\thanks{m.a.h.maccallum@qmul.ac.uk},
         G. Marcilhacy\inst{1}\thanks{gmarcilhacy@hotmail.com}
         \and  N.O. Santos\inst{1}\inst{2}\inst{3}\thanks{nilton.santos@upmc.fr}
         }

         \institute{LERMA-UPMC, Universit\'e Pierre et Marie Curie, Observatoire de Paris, CNRS, UMR 8112
       3 rue Galil\'ee, Ivry sur Seine 94200, France.
      \and
School of Mathematical Sciences,
  Queen Mary, University of London,
  Mile End Road, London E1 4NS, U.K.
\and Laborat\'orio Nacional de Computa\c{c}\~ao Cient\'{\i}fica,
25651-070 Petr\'opolis RJ, Brazil.}

\date{\today}

\begin{abstract}
{}{\ We re-examine the possibility that astrophysical jet collimation may
arise from the geometry of rotating black holes and the presence of
high-energy particles resulting from a Penrose process, without the help of
magnetic fields.}{Our analysis uses the Weyl coordinates, which are revealed
better adapted to the desired shape of the jets. We numerically integrate
the 2D-geodesics equations.}{We give a detailed study of these geodesics and
give several numerical examples.\ Among them are a set of perfectly
collimated geodesics with asymptotes $\rho =\rho _{1}$ parallel to the $z-$%
axis, with $\rho _{1}$ only depending on the ratios $\frac{\mathcal{Q}}{%
E^{2}-1}$ and $\frac{a}{M}$ , where $a$ and $M$ are the parameters of the
Kerr black hole, $E$ the particle energy and $\mathcal{Q}$ the Carter's
constant.}{} 
\end{abstract}

\maketitle

\institute{LERMA-UPMC, Universit\'e Pierre et Marie Curie, Observatoire de Paris, CNRS, UMR 8112
       3 rue Galil\'ee, Ivry sur Seine 94200, France.
      \and
School of Mathematical Sciences,
  Queen Mary, University of London,
  Mile End Road, London E1 4NS, U.K.
\and Laborat\'orio Nacional de Computa\c{c}\~ao Cient\'{\i}fica,
25651-070 Petr\'opolis RJ, Brazil.}



\section{Introduction}

It has long been speculated that a single mechanism might be at work in the
production and collimation of various very energetic observed jets, such as
those in gamma ray bursts (GRB) \cite{PirKumPan01,SheFraWhi03,Far03}, and
jets ejected from active galactic nuclei (AGN) \cite{SauTsiTru02} and from
microquasars \cite{MirRod94,MirRod99}. Here we limit ourselves to jets
produced by a black hole (BH) type core. The most often invoked process is
the Blandford-Znajek \cite{BlaZna77} or some closely similar mechanism
(e.g.\ \cite{PunCor90a,PunCor90,Pun01}) in the framework of
magnetohydrodynamics, always requiring a magnetic field. However such
mechanisms are limited to charged particles, and would be inefficient for
neutral particles (neutrons, neutrinos and photons), which are currently the
presumed antecedents of very thin and long duration GRB \cite{Far03}.
Moreover, even for charged particles, some questions persist (see for
instance the conclusion of \cite{Wil04}). Finally, while the observations of
synchrotron radiation prove the presence of magnetic fields, they do not
prove that those fields alone cause the collimation: magnetic mechanisms may
be only a part of a more unified mechanism for explaining the origin and
collimation of powerful jets, see \cite{Liv99} p.\ 234 and section 5, and,
in particular, for collimation of jets from AGN to subparsec scales, see
\cite{deZan00}.

Considering this background, it is worthwhile looking for other types of
model to explain the origin and structure of jets. Other models based on a
purely general relativistic origin for jets have been considered. A simple
model was obtained by \cite{OphSanWan96} by assuming the centres of galaxies
are described by a cylindrical rotating dust. That paper showed that
confinement occurs in the radial motion of test particles while the
particles are accelerated in the axial direction thus producing jets.
Another relativistic model was put forward in \cite{HerSan07}. This showed
that the sign of the proper acceleration of test particles near the axis of
symmetry of quasi-spherical objects and close to the horizon can change.
Such an outward acceleration, that can be very big, might cause the
production of jets.

However, these models show a powerful gravitational effect of repulsion only
near the axis, and are built in the framework of axisymmetric stationary
metrics which do not have an asymptotic behaviour compatible with possible
far away observations. So we want to explore the more realistic rotating
black hole, i.e.\ Kerr, metrics instead.

We thus address here the issue of whether it is possible, at least in
principle (i.e.\ theoretically) to obtain a very energetic and perfectly
collimated jet in a Kerr black hole spacetime without making use of magnetic
fields. Other authors (see \cite{BicSemHad93,Wil95,Wil04,deCar97} and
references therein) have made related studies to which we refer below. Most
such authors agree that the strong gravitational field generated by rotating
BHs is essential to understanding the origin of jets, or more precisely that
the jet originates from a Penrose-like process \cite{Pen69,Wil04} in the
ergosphere of the BH; collimation may also arise from the gravitational
field and that is the main topic in this paper.

Our work can therefore be considered as covering the whole class of models
in which particles coming from the ergosphere form a jet collimated by the
geometry. Although a complete model of an individual jet would require use
of detailed models of particle interactions inside the ergosphere, such as
that given by \cite{Wil04}, we 
show that thin and very long and energetic jets, with some generic features,
can be produced in this way. In particular the presence of a characteristic
radius, of the size of the ergosphere, around which one would find the most
energetic particles, might be observationally testable.

{}From a strictly general relativistic point of view, test particles in
vacuum (here, a Kerr spacetime) follow geodesics; this applies to both
charged and uncharged particles, although, of course, in an electrovacuum
spacetime, such as Kerr-Newman, charged particles would follow accelerated
trajectories, not geodesics. Thus, in Kerr fields, what produces an eventual
collimation for test particles, or not, is the form of the resulting
geodesics. Hence we discuss here the possibilities of forming an outgoing
jet of collimated geodesics followed by particles arising from a
Penrose-like process inside the ergosphere of a Kerr BH. We show that it is
possible in principle to obtain such a jet from a purely gravitational
model, but it would require the ``Penrose process'' to produce a suitable,
and rather special, distribution of outgoing particles.

The model is based on the following considerations.

Most studies of geodesics, e.g.\ \cite{Cha83}, employ generalized spherical,
i.e.\ Boyer-Lindquist, coordinates. We transform to Weyl coordinates, which
are generalized cylindrical coordinates, and are more appropriate, as we
shall see, for interpreting the collimated jets.

We consider test particles moving in the axisymmetric stationary
gravitational field produced by the Kerr spacetime, whose geodesic
equations, as projected into a meridional plane, are known \cite{Cha83}. Our
study is restricted to massive test particles, moving on timelike geodesics,
but of course massless test particles on null geodesics could be the subject
of a similar study. (Incidentally the compendium of \cite{Sha79} shows that
analytic studies of general timelike geodesics have been much less frequent
than detailed studies of more restricted problems.)

For particles outgoing from the ergosphere of the Kerr BH we examine their
asymptotic behaviour. Among the geodesic particles incoming to the
ergosphere, we discuss only the ones coming from infinity parallel to the
equatorial plane, because these are in practice the particles stemming from
the accretion disk. We show that only those with a small impact parameter
are of high enough energy to provide energetic outgoing particles.

In the ergosphere, a Penrose-like process can occur. In the original Penrose
process, an incoming particle decays into two parts inside the ergosphere.
It could also decay into more than two parts, or undergo a collision with
another particle in this region, or give rise to pair creation $%
(e^{-},e^{+}) $ from incident photons which would follow null geodesics. The
different possible cases do not affect our considerations, and that is why
we do not study them here, although the distribution function of outgoing
particles would be required in a more detailed model of the type discussed,
in particular to explain why only particles with low angular momentum and
not diverging from the rotation axis are produced. For detailed studies see
\cite{Wil95,Wil04,PirSha77}. After a decay, one (or more) of the particles
produced crosses the event horizon and irreversibly plunges into the BH,
while a second particle arising from the decay can be ejected out of the
ergosphere following a geodesic towards infinity. This outgoing particle
could be ejected so that asymptotically it runs parallel to the axis of
symmetry, but we do not discuss only such particles.

In our model there is no appeal to electromagnetic forces to explain the
ejection or the collimation of jets, though the particles therein may
themselves be charged. The gravitational field suffices, in the case of
strong fields in general relativity, which is the case near the Kerr BH,
provided the ergosphere produces particles of appropriate energy and initial
velocity. The gravitomagnetic part of the gravitational field then provides
the collimation. Hence, our model is, in this respect, simpler than the
standard model of \cite{BlaZna77}, and is in accordance with the analysis
given in \cite{Wil04}.

The paper starts with a study of Kerr geodesics in Weyl coordinates in
section \ref{Kerrgeod}; the next section studies the asymptotic behaviour of
geodesics of outgoing particles with $L_z=0$; section \ref{incoming}
analyses incoming particles stemming from the accretion;
a sample Penrose process and the plotting of geodesics are presented in
section \ref{penrose}; and finally we discuss in section \ref{jets} the
significance of our results for jets. In the conclusion, we succinctly
summarize our main results and evoke some perspectives.

\section{Kerr geodesics}

\label{Kerrgeod}

We start from the projection in a meridional plane $\phi=$ constant of the
Kerr geodesics in Boyer-Lindquist spherical coordinates $\bar{r}$, $\theta$
and $\phi$. 
The metric is
\begin{eqnarray}
\mathrm{d}s^2 & = & (\bar{r}^2+a^2\cos^2\theta) \left(\frac{\mathrm{d}\bar{r}%
^2}{\bar{r}^2-2M\bar{r}+a^2} + \mathrm{d}\theta^2\right)  \notag \\
&& +\frac{\sin^2 \theta}{ (\bar{r}^2+a^2\cos^2\theta)} \left(a\mathrm{d}t-(%
\bar{r}^2+a^2) \mathrm{d}\varphi \right)^2  \label{Kerr} \\
&& -\frac{(\bar{r}^2-2M\bar{r}+ a^2) }{(\bar{r}^2+a^2\cos^2\theta)} \left(%
\mathrm{d}t-a\sin ^2\theta \mathrm{d}\varphi\right)^2 .  \notag
\end{eqnarray}
where $M$ and $Ma$ are, respectively, the mass and the angular momentum of
the source, and we have taken units such that $c=1=G$ where $G$ is Newton's
constant of gravitation. The `radial' coordinate in (\ref{Kerr}) has been
named $\bar{r}$ because it is more convenient for us to use the rescaled
coordinate $r=\bar{r}/M$ \cite{ONe95}. The projected timelike geodesic
equations are then
\begin{eqnarray}
M^2\dot{r}^2&=& \frac{(a_4r^4+a_3r^3+a_2r^2+a_1r+a_0)}{ \left[r^2+\left(%
\frac{a}{M}\right)^2\cos^2\theta\right]^{2}},  \label{1} \\
M^2\dot{\theta}^2&=& \frac{b_4\cos^4\theta+b_2\cos^2\theta+b_0}{%
(1-\cos^2\theta) \left[r^2+\left(\frac{a}{M}\right)^2\cos^2\theta\right]^{2}}%
,  \label{2}
\end{eqnarray}
with coefficients
\begin{eqnarray}
a_0&=&-\frac{a^2\mathcal{Q}}{M^4},  \label{3} \\
a_1&=&\frac{2}{M^2}\left[(aE-L_z)^2+\mathcal{Q}\right],  \label{4} \\
a_2&=&\frac{1}{M^2}\left[a^2(E^2-1)-L_z^2-\mathcal{Q}\right],  \label{4a} \\
a_3&=&2,  \label{5} \\
a_4&=& E^2 -1,  \label{6}
\end{eqnarray}
and
\begin{eqnarray}
b_0&=&\frac{\mathcal{Q}}{M^2},  \label{7} \\
b_2&=&\frac{1}{M^2}\left[a^2(E^2-1)-L_z^2-\mathcal{Q}\right]=a_2,  \label{8}
\\
b_4&=&-\left(\frac{a}{M}\right)^2(E^2-1);  \label{9}
\end{eqnarray}
where the dot stands for differentiation with respect to an affine parameter
and $E$, $L_z$ and $\mathcal{Q}$ are constants. Here Chandrasekhar's $%
\delta_1$ has been set to 1, its value for timelike curves. Assuming that
the affine parameter is proper time $\tau$ along the geodesics, then these
equations implicitly assume a unit mass for the test particle \footnote{%
An alternative interpretation is to assume that for a particle of mass $m$,
the affine parameter $\tau/m$ has been used \cite{Wil72,Wil95}.}, so that $E
$ and $L_z$ have the usual significance of total energy and angular momentum
about the $z$-axis, and $\mathcal{Q}$ is the corresponding Carter constant
(which, as described in \cite{HugPenSom72} for example, arises from a
Killing tensor of the metric, while $E$ and $L_z$ arise from Killing
vectors). With this understanding, $E$, $L_z$, and $\mathcal{Q}$ have the
dimensions of Mass, Mass$^2$ and Mass$^4$ respectively, in geometrized
units, while $\delta_1$, though 1 numerically, has dimensions Mass$^2$, as
do all the $a_i$ and $b_i$. In this paper we consider only particles on
unbound geodesics with $E\geq 1$. (For the conditions for existence of a
turning point, giving bound geodesics, which are related to the parameter
values for associated circular orbits, see \cite{Cha83,Wil95,Wil04}.)

The dimensionless Weyl cylindrical coordinates, in multiples of geometrical
units of mass $M$, are given by
\begin{equation}
\rho =\left[ (r-1)^{2}-A\right] ^{1/2}\sin \theta ,\;\;z=(r-1)\cos \theta ,
\label{10}
\end{equation}%
where
\begin{equation}
A=1-\left( \frac{a}{M}\right) ^{2}.  \label{11}
\end{equation}%
{}From (\ref{10}) we have the inverse transformation
\begin{eqnarray}
r &=&\alpha +1,  \label{11a} \\
\sin \theta &=&\frac{\rho }{(\alpha ^{2}-A)^{1/2}},\;\;\cos \theta =\frac{z}{%
\alpha },  \label{12}
\end{eqnarray}%
with
\begin{equation}
\alpha =\frac{1}{2}\left( [\rho ^{2}+(z+\sqrt{A})^{2}]^{1/2} +[\rho ^{2}+(z-%
\sqrt{A})^{2}]^{1/2} \right).  \label{13}
\end{equation}
Here we have assumed $A\geq 0$, and taken the root of the second degree
equation obtained from (\ref{12}) for the function $\alpha ^{2}(\rho ,z)$
that allows the extreme black hole limit $A=0$. The other root, in this
limit, is the constant $\alpha =0$.

The equation (\ref{13}) shows that in the $(\rho,\,z)$ plane the curves of
constant $\alpha$ (constant $r$) are ellipses with semi-major axis $\alpha$
and eccentricity $e=\sqrt{A}/\alpha$: for large $\alpha$, these approximate
circles. Note that $\rho=0$ consists of the rotation axis $\theta =0$ or $\pi
$ together with the ergosphere surface.

Now, with (\ref{11a}) and (\ref{12}) we can write the geodesics (\ref{1})
and (\ref{2}) in terms of $\rho $ and $z$ coordinates, producing the
following autonomous system of first order equations
\begin{eqnarray}
M\dot{\rho} &=&\frac{\displaystyle{\frac{P\alpha ^{3}\rho }{\alpha ^{2}-A}+
\frac{S(\alpha ^{2}-A)z}{\alpha \rho }}}{\displaystyle{(\alpha +1)^{2}\alpha
^{2}+\left( \frac{a}{M}\right) ^{2}z^{2}}},  \label{14} \\
M\dot{z} &=&(Pz-S)\alpha \left[ (\alpha +1)^{2}\alpha ^{2}+\left( \frac{a}{M}
\right) ^{2}z^{2}\right] ^{-1},  \label{15}
\end{eqnarray}
where
\begin{eqnarray}
P &=&\epsilon_1\left[ a_{4}(\alpha+1)^{4}+a_{3}(\alpha +1)^{3}\right.
\label{16} \\
&&\left. \phantom{a_{4}(\alpha)^{4}}+a_{2}(\alpha +1)^{2}+a_{1}(\alpha
+1)+a_{0}\right] ^{1/2},  \notag \\
S &=&\epsilon_1\epsilon _{2}(b_{4}z^{4}+b_{2}\alpha ^{2}z^{2}+b_{0}\alpha
^{4})^{1/2},  \label{17}
\end{eqnarray}
and $\epsilon _{i}=\pm 1$ for $i=1,2$: $\epsilon _1$ indicates whether the
geodesic is incoming or outgoing in $r$ (i.e.\ the sign of $\dot{r}$), while
$\epsilon _1\epsilon _2$ indicates whether $\theta$ is increasing or
decreasing. Note we always mean the non-negative square roots to be taken.

The ratio between the first order differential equations (\ref{14}) and (\ref%
{15}) yields the special characteristic equation of this system of equations
\begin{equation}
\frac{dz}{d\rho }=\frac{(|P|z-\epsilon_2|S|)(\alpha ^{2}-A)\alpha ^{2}\rho }{%
|P|\alpha^{4}\rho ^{2}+\epsilon_2|S|(\alpha ^{2}-A)^{2}z}.  \label{18}
\end{equation}

We restrict our study to the quadrant $\rho >0$ and $z>0$ in the projected
meridional plane (orbits in fact spiral round in $\phi$ in general: this
information is contained in the conserved $L_z$). The results for the other
three quadrants will follow by symmetry, although this symmetry does not
imply that individual geodesics are symmetric with respect to the equatorial
plane. From numerical solutions of the geodesics we obtained asymmetrical
geodesics, confirming the analysis in \cite{Wil95,Wil04}. Geodesics can also
cross the polar axis, which would be represented by a reflection from $\rho=0
$ back into the quadrant.

Geodesics going to or coming from the expected accretion disk would, if the
disk were thin, go to or from values of $\rho$ much larger than $z$. In this
limit ($\rho \gg z$ and $\rho \gg \sqrt{A}$) , we have
\begin{eqnarray}
\alpha &=& \rho(1+O(\rho ^{-2})), \\
|P|&=&\sqrt{a_4}\rho^2(1+k/\rho + O(\rho ^{-2})), \\
&& \mathrm{where}~~ k=\frac{2E^2-1}{E^2-1}, \\
|S|&=& \sqrt{b_0}\rho^2(1+O(\rho ^{-2})),
\end{eqnarray}
and thus
\begin{equation}
\frac{dz}{d\rho }\approx \frac{z-\epsilon_2 z_{1}+zk/\rho}{\rho(1+k/\rho) }%
+O(\rho ^{-3}),  \label{19a}
\end{equation}
where
\begin{equation}
z_{1}=\left( \frac{b_{0}}{a_{4}}\right) ^{1/2}=\frac{1}{M}\left( \frac{%
\mathcal{Q}}{E^{2}-1}\right) ^{1/2},  \label{20}
\end{equation}%
and we have to assume $\mathcal{Q}\geq 0$ to obtain a real $z _{1}$ (the
form of (\ref{2}) makes it obvious that geodesics with $\mathcal{Q}<0$ are
bounded away from the equatorial plane $\cos \theta =0$, though those with
very small $|\mathcal{Q}|$ could still satisfy $\rho \gg z$).

The truncated series development of $\mbox{\rm d} z/\mbox{\rm d}\rho$ now
yields
\begin{equation}
\frac{dz}{d\rho }\approx \frac{1}{\rho }\left( z-\epsilon_2 z_{1}+\frac{%
\epsilon_2 k z_{1}}{\rho}\right)+O(\rho ^{-3}).  \label{19}
\end{equation}
If $\epsilon _{2}=-1$ the curve crosses the equatorial plane and gives
similar asymptotic behaviour in the second quadrant, so we can take $%
\epsilon _{2}=1$.

A thicker accretion disk would absorb or release particles on geodesics with
larger values of $z/\rho$, which might include particles with $\mathcal{Q}<0$%
.

Geodesics in an axial jet would have $z\gg \rho$. For this limit, we first
observe that from (\ref{17}) we have
\begin{equation}
S\approx \epsilon _{2}\left[ (b_{0}+b_{2}+b_{4})z^{4}+(2b_{0}+b_{2})\rho
^{2}z^{2}\right] ^{1/2}+O(z^{-1}),  \label{21}
\end{equation}%
where
\begin{eqnarray}
b_{0}+b_{2}+b_{4} &=&-\left( \frac{L_{z}}{M}\right) ^{2}\leq 0,  \label{22a}
\\
2b_{0}+b_{2} &=&\frac{1}{M^{2}}\left[ a^{2}(E^{2}-1)-L_{z}^{2}+\mathcal{Q}%
\right] .  \label{22b}
\end{eqnarray}%
Hence in this limit $S$ is well defined and real for indefinitely small $%
\rho /z$ only for $L_{z}=0$. The geodesics obeying this restriction, imposed
after similar reasoning, were studied by \cite{BicSemHad93}, but in BL or
Kerr-Schild coordinates. Here we re-examine these geodesics in the more
revealing cylindrical coordinates.

Before doing so, we may note that in contrast to geodesics with $L_z \neq 0$%
, geodesics with $E^2>1$ and $L_z=0$ may lie arbitrarily close to the polar
axis \cite{Car68}. For $L_z \neq 0$, the value of $S^2$ at the axis is $%
-z^2L_z^2 <0$ which is not allowed and thus there is some upper bound $%
\theta_0$ on $\theta$. The value of $S^2$ at $\theta=\pi/2$ is $b_0\alpha^2$
so if $\mathcal{Q}<0$ there is also a lower bound $\theta_1$ on $\theta$.

\section{Geodesics with $L_z=0$}

\label{L0geods}

We shall discuss unbounded ($E^2>1$) outgoing geodesics. Corresponding
incoming geodesics will follow the same curves in the opposite direction.

For $L_z=0$, $S^2$ factorizes as
\begin{equation}
S^2 = (\alpha^2-z^2)({\mathcal{Q}}\alpha^2 + a^2(E^2-1)z^2)/M^2.
\label{Sfactors}
\end{equation}
Hence $S$ can only be zero at the symmetry axis, where $\cos\theta=1$, $%
\alpha=z$ and $\rho=0$, or, if ${\mathcal{Q}}<0$, at some $z/\alpha=(|{%
\mathcal{Q}}|/a^2[E^2-1])^{1/2} = \cos\theta_1$, say. Correspondingly $\dot{%
\theta}=0$ only at the axis, at $\theta=\theta_1$ if ${\mathcal{Q}}<0$, and
at $\alpha \rightarrow \infty$.

Thus for $L_{z}=0 $ and $\mathcal{Q}<0$, geodesics which initially have $%
\dot{\theta}<0$ will become asymptotic to $\theta=\theta_1$.
The angle may be narrow if
\begin{equation}
a^{2}(E^{2}-1 )-|\mathcal{Q}|\ll |\mathcal{Q}|,  \label{39}
\end{equation}
and then $\theta_1 \ll 1$. Such geodesics may provide a conical jet, as
discussed later.

Our other polynomial, $P^2$, can be written as
\begin{eqnarray}
P^2 &=& (E^2-1)\left(r^4+\frac{a^2}{M^2}r^2+\frac{2a^2}{M^2}r\right)
\label{formofP} \\
&& +2r^3+2\frac{a^2}{M^2}r- \frac{\mathcal{Q}}{M^2} \left(r^2-2r+\frac{a^2}{%
M^2}\right).  \notag
\end{eqnarray}
>From this form it easily follows that any unbound geodesic ($E^2>1$) with $%
L_{z}=0$ has at most one turning point in $r$ (i.e.\ value such that $\dot{r}
=0$) and this, if it exists, lies inside the horizon (and a fortiori inside
the ergosphere) \cite{SteWal74}. The argument is very simple. If $E^{2}>1$
and ${\mathcal{Q}}\leq 0$, then $P^2$ is strictly positive for all $r>0$. If
${\mathcal{Q}}>0$, $P^2$ is negative at $r=0$ but positive at the outer
black hole horizon (where $r^2-2r+a^2/M^2 =0$), so its one zero lies inside
the black hole. This implies that unbounded outgoing geodesics followed by
particles with $L_z=0$ must come from the ergosphere. Correspondingly,
geodesics incoming from infinity with $L_z=0$ will fall into the ergosphere.

Although there are no turning points of $r$, one can have turning points of $%
\rho$, if $\epsilon_2=-1$. Such turning points are solutions of the equation
\begin{equation}
D(\rho,\,z)\equiv |P|\alpha^{4}\rho ^{2}-|S|(\alpha ^{2}-A)^{2}z=0
\label{Drhoz}
\end{equation}
where $D(\rho ,\,z)$ is the denominator of (\ref{18}) with $\epsilon_2=-1$.
At each of these turning points $(\rho_{2}$, $z_{2})$, $dz/d\rho \rightarrow
\infty $, which means that the geodesics have a vertical tangent (parallel
to the $z$-axis). Before reaching the turning point, these geodesics have $%
\mbox{\rm d} \rho/\mbox{\rm d} z>0$, and, at any $z$, $\mbox{\rm d} z/%
\mbox{\rm d} \rho > \mbox{\rm d} z/\mbox{\rm d}\rho|_D$, where $\mbox{\rm d}
z/\mbox{\rm d}\rho|_D$ is the slope of the curve (\ref{Drhoz}), and
afterwards they have $\mbox{\rm d} \rho/\mbox{\rm d} z<0$, implying that
they subsequently cross the axis. They will then cross the curve (\ref{Drhoz}%
) again but from above in the $(\rho,\,z)$ plane and hence with $\mbox{\rm d}
z/\mbox{\rm d} \rho < \mbox{\rm d} z/\mbox{\rm d}\rho|_D$, and afterwards
stay in the region outside (\ref{Drhoz}).

For outgoing geodesics outside (\ref{Drhoz}) which reach points at large $z$
and $\rho$ ($\gg \sqrt{A}$), then unless the ratio of $z$ to $\rho$ is very
large (the case which we discuss next) or very small, approximating (\ref{18}%
) gives $\mbox{\rm d} z/\mbox{\rm d}\rho \approx z/\rho$, so all such
geodesics approximate $\rho=Cz$ for suitable $C$, regardless of the sign of $%
\epsilon_2$.

In the limit $z\gg \rho$ and $z \gg \sqrt{A}$,
\begin{eqnarray}
\alpha &=& z(1+O(z ^{-2})), \\
|P|&=&\sqrt{a_4}\,z^2(1+k/z + O(z ^{-2})), \\
|S|&=&\sqrt{2b_0+b_2}\;\rho z(1+O(z ^{-2})),
\end{eqnarray}
so the equation (\ref{18}) can be approximated by
\begin{equation}
\frac{dz}{d\rho } = \frac{z(1+k/z)}{\rho(1+k/z)+\epsilon_2\rho_1} +O\left(%
\frac{1}{\rho z}\right),  \label{largez}
\end{equation}
where
\begin{eqnarray}
\rho _{1} &=&\left( \frac{2b_{0}+b_{2}}{a_{4}}\right) ^{1/2}=\rho _{e}\left[
1+\frac{\mathcal{Q}}{a^{2}(E^{2}-1)}\right] ^{1/2},  \label{30}
\end{eqnarray}
and $\rho _{e}\equiv a/M$. Here $\rho_1$ is real if $a^{2}(E^{2}-1)+\mathcal{%
Q}>0$ but we see from (\ref{Sfactors}) that for $S$ to be real near the
axis, this condition must be satisfied.

\begin{figure}[ht]
\centering
\includegraphics[width=8cm]{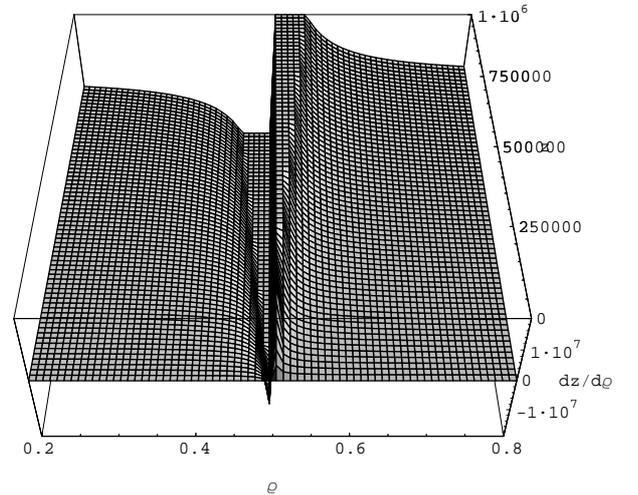}
\caption{Plot of the surface $\mbox{\rm d} z/\mbox{\rm d} \protect\rho =f(%
\protect\rho ,z)$ given by equation (\protect\ref{18}) for an outgoing
particle with the parameters $E=10^4$, $L_z=0$ and $\mathcal{Q}= -9\times
10^3$ for a black hole with parameters $M=1$ and $a=0.5$, in the case where $%
\protect\epsilon_2=-1$. The points where $dz/d\protect\rho \rightarrow \pm
\infty$ correspond to the asymptotes given by the equation (\protect\ref{30}%
), $\protect\rho=\protect\rho_1=0.49991$.}
\label{eps2minus}
\end{figure}

In Figure \ref{eps2minus} we show a plot of the values of $\mbox{\rm d} z/%
\mbox{\rm d}\rho$, using (\ref{18}), for $\varepsilon _{2}=-1$, with the
parameters $M=1$, $a=1/2$, $E=10^{4}$, $\mathcal{Q}=-9\times 10^{3}$. The
only asymptotes are parallel to the $z$ axis at $\rho =\rho _{1}$ as
expected from (\ref{30}).

\begin{figure}[ht]
\centering\includegraphics[width=6cm,height=6cm]{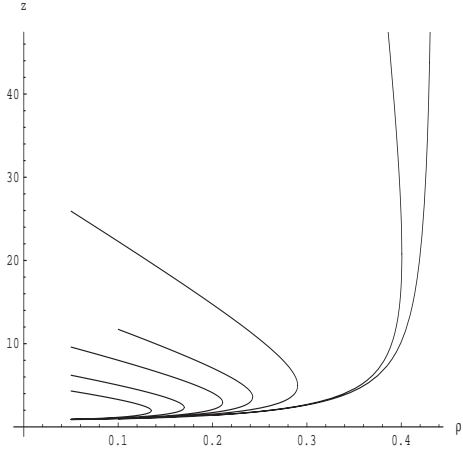}
\caption{Plots of geodesics obeying the equation (\protect\ref{18}), showing
the turning points. From left to right these curves start at $\protect\rho%
_0=0.07$ and $z_0=0.98$, 0.95, 0.93, 0.92, 0.91, 0.9 and 0.89935501}
\label{fig3}
\end{figure}

We also plot in Figure \ref{fig3} a set of such outgoing geodesics obeying (%
\ref{18}), for the same values of the parameters of the BH ($a=1/2$, $M=1$)
and of the particle ($L_{z}=0$, $\mathcal{Q}=-2.2\times 10^5$, $E=2.10^3$,
so $\rho_1=0.441588$), but with different initial values of the position.
The set of turning points of these geodesics is the curve defined by
equation (\ref{Drhoz}). For the rightmost of these geodesics, the numerical
integration was also continued back towards the ergosphere as far as $\rho =
10^{-4}$, $z=0.843407 < \sqrt{A} = 0.866025$.

To confirm the picture obtained from these numerical experiments, one can
show, without assuming $z \gg \rho$, the existence of exactly one zero of $D$
on any curve $r=$ constant, $0<\theta<\pi/2$, so that the conclusion that a
geodesic has at most one turning point in $\rho$ is not an artefact of the
approximation at large $z$. The argument is as follows.

Along an $r=$ constant curve, $|P|$ and $\alpha$ are constant, $\rho =
\rho_0 \sin \theta$ and $z = \alpha \cos \theta$, where $\rho_0$ is a
constant (related to $r$). Then
\begin{eqnarray}
D &=& |P|\alpha^4\rho_0^2 \sin^2 \theta-|S|(\alpha^2-A)^2 \alpha \cos \theta
\end{eqnarray}
where, defining
\begin{eqnarray}
F&\equiv &(b_0+ |b_4|\cos^2\theta)^{1/2}  \notag \\
&=&[(\mathcal{Q} +a^2(E^2-1)\cos^2 \theta)]^{1/2}/M,
\end{eqnarray}
from (\ref{Sfactors}) we have $|S| = \alpha^2 F \sin \theta $. For real $S$
we need $F \geq 0$ and as $\theta$ decreases, $F$ increases.

At $\theta = \pi/2$, $D>0$, while at small $\theta$, $D<0$. Hence there is
at least one zero of $D$. Let the largest one be at $\theta=\theta_0$ say.
On the $r=$ constant curve, we will then have
\begin{equation}
D= \alpha^3 (\alpha^2-A)^2 \frac{\sin \theta}{\sin \theta_0} (F_0 \sin
\theta \cos \theta_0 - F \cos \theta \sin \theta_0).
\end{equation}
Here we have used $D=0$ at $\theta=\theta_0$ to substitute for the constant $%
|P|\alpha^4 \rho_0^2 \sin^2 \theta_0$ in terms of $\alpha$ and $\theta_0$.
As $\theta$ decreases from $\theta_0$, $\sin \theta$ decreases, $\cos \theta$
increases and $F$ increases. Thus the combination $(F_0 \sin \theta \cos
\theta_0 - F \cos \theta \sin \theta_0)$ becomes and stays negative for all $%
\theta < \theta_0$. Thus $D<0$ for all $\theta < \theta_0$, though $D$
approaches 0, due to the further factor $\sin \theta$, as $\theta
\rightarrow 0$. This implies that the only points on $r=$ constant such that
$D=0$ are at $\theta=0$ and $\theta=\theta_0$.

For large $z$ we see from (\ref{largez}) that the turning points lie
approximately on a curve $\rho=\rho_1 z/(z+k)$ or $z=-k\rho/(\rho-\rho_1)$.
Actually, the differential equation for large $z$, if we drop the $1/z^2$
terms, has an analytic solution
\begin{equation}
Ckz = k\rho + \rho_1 z \ln [z/(z+k)],  \label{approxz}
\end{equation}
where $C$ is a constant of integration, so
\begin{equation}
\rho = \rho_1 z/k \ln (1+k/z) + Cz \rightarrow \rho_1 + Cz \ldots
\label{lgzapprox}
\end{equation}
as $z \rightarrow \infty$. Keeping the next order terms in $1/z$ would be
inconsistent with the terms dropped during the derivation. (Similarly, there
is an analytic solution for the approximate equation (\ref{19a}) at large $%
\rho$ which gives $z = z_1 \rho \ln(1+ \rho/k)/k + c \rho$ with similar
interpretation.)

>From (\ref{lgzapprox}), either (a) $\rho/z$ is approximately constant or (b)
$\rho \rightarrow \rho_1$. In case (a), we note that for consistency of the
approximation $z \gg \rho$, $C$ must be small, although the conclusion is
the same as was reached above merely with the assumption that both $z$ and $%
\rho$ are $\gg \sqrt{A}$. In case (b), we have a limit-outgoing geodesic for
which $\rho <\rho _{1}$ at all points and as $z\rightarrow \infty $, $\rho
\rightarrow \rho _{1}$. This limit is obtained since the turning point for $%
\rho$ has $z_{2}\rightarrow \infty$ when $\rho _{2}\rightarrow \rho _{1}$.
We can see from (\ref{approxz}) that the coordinate $z_{2}$ of this turning
point tends to infinity like $z_{3}= -k\rho_1/(\rho_2-\rho_1)$. The
geodesics asymptotic to $\rho _{1}$ would provide a perfectly collimated jet
parallel to $z$.

One might think (and we initially thought) that there also existed geodesics
eventually tending to the same asymptote but approaching it from the right
in the $(\rho,\,z)$ plane (for example, directly from the accretion disk, or
coming from the ergosphere but with a turning point $\rho _{2}>\rho _{1}$).
However, such geodesics do not exist, since they require that $\mbox{\rm d}
z/ \mbox{\rm d}\rho <0$ in the limit $z\gg \rho $ and for $\rho >\rho _{1}$,
contradicting (\ref{largez}) which implies $\mbox{\rm d} z/\mbox{\rm d}\rho
>0$. This is entirely in agreement with the results of Stewart and Walker
\cite{SteWal74}.

The geodesics in $\rho >\rho _{1}$ may asymptote to any ratio $z/\rho$, from
(\ref{largez}). Moreover, geodesics which do turn in $\rho$ then cross the
axis, cannot cross the curve $D=0$ from below again, and so cross it from
above and also asymptotically have some fixed ratio $z/\rho$.

For astrophysical applications, it may be important to write the results in
the normal units of length and time. We have, from (\ref{14}), that
asymptotically for outgoing particles in $z\gg \rho $, $\dot{\rho}%
\rightarrow 0$, $\dot{t}\rightarrow E$ , $\overset{\cdot }{z}$ $>0$ (for $%
-\epsilon _{2}=1=\epsilon _{1}$) and (\ref{15}) is given by
\begin{equation}
M\dot{z}\approx \sqrt{a_{4}};  \label{26b}
\end{equation}%
hence, restoring normal units of length and time and taking a particle of
mass $m$, the asymptotic value $v$ of the speed of outgoing particles is
given by
\begin{equation}
mM\dot{z}\approx \left( \frac{E^{2}}{c^{4}}-m^{2}\right) ^{1/2}=m\gamma
\frac{v}{c},  \label{26c}
\end{equation}%
where we have used (\ref{6}) and
\begin{equation}
\gamma =\left[ 1-\left( \frac{v}{c}\right) ^{2}\right] ^{-1/2}=\frac{E}{%
mc^{2}},  \label{26d}
\end{equation}%
is the Lorentz factor. Hence, asymptotically, the speed of the particle is
\begin{equation}
v=c\frac{M\dot{z}}{\dot{t}}=c\left( 1-\frac{m^{2}c^{4}}{E^{2}}\right) ^{1/2},
\label{26e}
\end{equation}%
which is ultrarelativistic if $E\gg \sqrt{mc^{2}}$. From (\ref{26c}) we have
that asymptotically limit-outgoing particles have an uniform motion parallel
to the $z$ axis.

\section{Incoming particles}

\label{incoming}

We describe as ``incoming particles" the particles, with parameters $%
E^{\prime }$, $L_{z}^{\prime }$, and $\mathcal{Q}^{\prime }$, coming into
the ergosphere following unbound geodesics and having a turning point in $z$
(i.e.\ such that $\overset{\cdot }{z}$ $=$ $0$). Such turning points \{$\rho
_{4}$, $z_{4}$\} are defined as solutions of the equation%
\begin{equation}
N_{2}(\rho _{4},z_{4})=0
\end{equation}
where%
\begin{equation}
N_{2}=Pz-S
\end{equation}
is the relevant factor in the numerator of the right side of (\ref{18}). As
remarked earlier we need only consider $\epsilon _{2}=1$. For large $\rho$
these turning points are approximately at $z=z_1(1-k/\rho)$, from (\ref{19}%
). The set of these turning points \{$\rho _{4},z_{4}$\} forms a curve of
the points of each geodesic with horizontal tangent (i.e.\ parallel to the $%
\rho - $axis). Each geodesic has a positive slope $dz/d\rho >0$ for $\rho
<\rho _{4} $, a negative one for $\rho >\rho _{4}$ and a maximum at the
turning point. There exists a limit-incoming geodesic with its turning point
at the infinite $\rho _{4}\rightarrow \infty $ for $z_{4}\rightarrow z_{1}$,
i.e. with the asymptote $z=z_{1}$, when $\rho \rightarrow \infty $.\newline

\begin{figure}[htbp]
\centering\includegraphics[width=8cm]{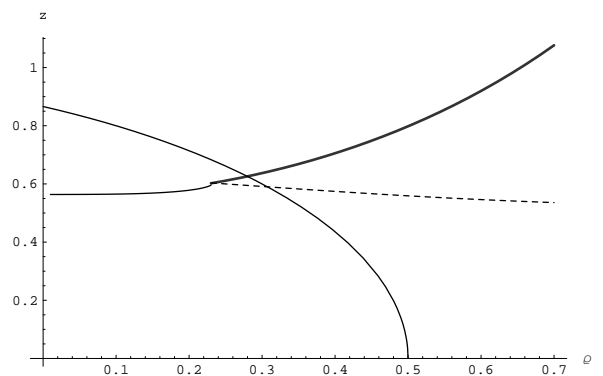}
\caption{Penrose process. Plots of the ingoing particle (dashed line) coming
asymptotically, for $\protect\rho\rightarrow\infty$, from $z_1=1.58116$, to
the turning point ($\protect\rho=0.23$, $z=0.59$) located inside the
ergosphere, whose boundary is indicated, where it decays into an infalling
particle and an outgoing particle (bold line) following a geodesic
asymptotic to $\protect\rho=\protect\rho_1=1.64341$ when $z \rightarrow
\infty$. The parameters of the black hole are $M=1 $ and $a=0.5$. The
parameters of the three geodesics are $E_{in}=200$, $E_{out}=202$, $%
E_{fall}=-2$, $L_{z,in}=L_{z,fall}=-100$, $L_{z,out}=0$, ${\mathcal{Q}}={%
\mathcal{Q}}^{\prime }= 100000.$}
\label{fig4}
\end{figure}

A test particle with parameters $E^{\prime }$, $L_{z}^{\prime }$, and $%
\mathcal{Q}^{\prime }$ coming from infinity (in practice from the accretion
disk) parallel to $z=0$ towards the axis of the black hole corresponds to a
geodesic which, in the limit $\rho \rightarrow \infty $, has an asymptote
defined by $z=z_{1}=$ constant where $z_{1}$ is the impact parameter.
Therefore it is a limit-incoming particle with $z<z_{1}$, $\dot{z}<0$, $\dot{%
\rho}<0$, $\epsilon _{1}=\epsilon _{2}=1$. In the limit $\rho \gg z$, $%
dz/d\rho =0$, so the tangent has to be parallel to the $\rho $ axis and (\ref%
{20}) produces
\begin{equation}
z_{1}=\frac{1}{M}\left( \frac{\mathcal{Q}^{\prime }}{E^{\prime 2}-1}\right)
^{1/2}.  \label{29}
\end{equation}%
We have plotted in Fig.\ \ref{fig4} (see section \ref{penrose}) an example
of a geodesic of an incoming particle. We see that, unlike $\rho _{1}$, $%
z_{1}$ does not depend upon the black hole parameter $a$. The incoming
particles come from the accretion disk, which means that their energy $%
E^{\prime }$ is rarely very big, which means \textit{rarely as big as $a$ or
$M$}, which characterize the black hole energy, but instead of the same
order as 1 or $\sqrt{\mathcal{Q}^{\prime }}$. When $E^{\prime }\rightarrow 1$
then $z_{1}\rightarrow \infty $ if $\mathcal{Q}^{\prime }\neq 0$. However,
if $E^{\prime }$ is very big compared to 1 and $\sqrt[4]{\mathcal{Q}^{\prime
}}$, then $z_{1}\rightarrow 0$.

For a given $\mathcal{Q}^{\prime}$, the most energetic incoming particles
are those with a small impact parameter $z_1$, near to zero. Hence only a
thin slice of the accretion disk can participate with the greatest
efficiency in producing Penrose processes leading to the most intense
possible jet. The point where the ergosphere surface intersects the $z$ axis
is $z_e=\sqrt{A}$. The value of $z_e$, for the incoming particles, does not
play a role like that of $\rho_e$ for the outgoing particles (compare (\ref%
{30}) to (\ref{29})).

\section{Penrose process and plotting of geodesics}

\label{penrose}

To make a jet using the geodesics discussed above, we would have to assume
that incoming particles arrive in the ergosphere and undergo a Penrose
process. As mentioned earlier, in its original version \cite{Pen69}, each
particle may be decomposed into two subparticles and one of them may cross
the horizon and fall irreversibly into the BH, while the other is ejected to
the exterior of the ergosphere; or the incoming particle may collide with
another particle resulting in one plunging into the BH and the other being
ejected to the exterior. The second case can correspond to a creation of
particles, say $e^{-}$ and $e^{+}$ from an incoming photon ($\delta =0$)
interacting with another inside the ergosphere. We do not present here all
the possible cases, which are exhaustively studied, especially for AGN, in
\cite{Wil95,Wil04}. There is also observational evidence for a close
correlation between the disappearance of the unstable inner accretion disk
and some subsequent ejections from microquasars such as GRS 1915+1105 \cite%
{MirRod94,MirRod99}, which from our point of view could correspond to the
instability causing disk material to fall through the ergosphere and to then
give rise to a burst of ejecta from Penrose-like processes. Here we are
mainly interested in the outgoing particles which follow geodesics that tend
asymptotically towards a parallel to the $z$ axis, as described in the
earlier section \ref{L0geods}. These events are closely dependent on the
possibilities allowed by the conservation equations. In the case when the
incoming particle splits into two \cite{ReeRufWhe76}, the conservation
equations of the energy and angular momentum are
\begin{eqnarray}
E_{in} &=&E_{out}+E_{fall},  \label{40} \\
L_{z,in} &=&L_{z,out}+L_{z,fall}.  \label{41}
\end{eqnarray}
We know from (\ref{21}) that for the outgoing particles we study $%
L_{z,out}=0 $. The particles falling irreversibly into the BH have energy $%
E_{fall}$ and angular momentum $L_{z,fall}$. The energy and the angular
momentum of a particle in the ergoregion can be negative which is the basis
of the Penrose process. The particle falling into the black hole has
negative energy, and hence the outgoing particle, leaving the ergosphere,
has a bigger energy than the incoming particle,
\begin{equation}
E_{out}=E_{in}+|E_{fall}|.  \label{42}
\end{equation}

We have plotted numerically the geodesics for incoming, outgoing and falling
particles with the following values for the parameters: $a/M=1/2$, $%
E_{in}=200$, $E_{out}=202$, $E_{fall}=-2$, $L_{z,in}=L_{z,fall}=-100$ and $%
\mathcal{Q}=\mathcal{Q}^{\prime }=10^{5}$. These values are chosen in such a
way that the geodesics meet inside the ergoregion situated in the interior
of the ergosphere
\begin{equation}
z^{2}=\left\{ 1-\rho _{e}^{2}\left[ 1-\left( \frac{\rho }{\rho _{e}}\right) %
\right] \right\} \left[ 1-\left( \frac{\rho }{\rho _{e}}\right) \right] ,
\end{equation}
and they produce for the asymptotes of the outgoing particles $\rho
_{1}=1.64341$ and for the incoming $z_{1}=1.58116$. The curves are built
with the initial conditions $z[1.55]=20$ for outgoing particles and $%
z[0.25]=0.60$ for incoming particles. Their intersection is situated at the
point $\rho _{i}=0.23$ and $z_{i}=0.59$ inside the ergosphere which we can
take as the initial condition for the falling particle and from where we
trace the three curves (see fig.\ 4).

The exhibition of these numerical solutions with an outgoing geodesic which
leaves the ergosphere after the Penrose process and has vertical asymptote
with the value $\rho_1$ precisely equal to (\ref{30}) confirms that a model
based on such geodesics is possible.

\section{Implications for jet formation}

\label{jets}

We have shown that to obtain a jet of particles close to the rotation axis,
it must be formed from particles with (almost) zero angular momentum, $L_z=0$%
. If we consider only particles with $L_z=0$, there is among them a subset
which give a perfectly collimated jet, i.e.\ a set of geodesics exactly
parallel to the axis: for each allowed value of $\mathcal{Q}$ they form a
ring of radius
\begin{equation}
\rho _{1}= \rho _{e}\left[ 1+\frac{\mathcal{Q}}{a^{2}(E^{2}-1)}\right]
^{1/2},
\end{equation}
whwre $\rho_e=a/M$. Note that $\rho =\rho_e$ is the circumference of the
ergosphere in the equatorial plane. For large $E$ the set of all these
geodesics will give a jet of radius of the order of $\rho_e$, with a density
of particles dependent on the distribution in $\mathcal{Q}$ for given $E$,
which remains perfectly collimated all the way to infinity.

We note that all other geodesics with $L_z=0$ will spread out from the axis
along lines $z=K\rho$. An astrophysical jet will of course be of only finite
extent and not perfectly collimated, so it could include such geodesics for
suitably large $K$, as well as geodesics with a small $L_z \neq 0$.

Thus forming a collimated jet of particles from a Penrose-like process, this
jet having a narrow opening angle, for a rotating black hole without an
electromagnetic field, depends on the initial distribution of particles
leaving the ergosphere, or of some non-gravitational collimating force, even
if we consider only particles with $L_z=0$.

On the other hand, outgoing particles with small energies, namely of the
order of their rest energy, $E\approx 1$, and $\mathcal{Q} >0$ have
asymptotes parallel to the $z$ axis with $\rho_1\gg\rho_e$.

This predicted scale of the region of confined highly energetic particles
might provide a test if the accretion disk parameters provided values for
the BH mass and angular momentum, in a manner such as discussed in \cite%
{McCShaNar06} and papers cited therein, and if the transverse linear scale
of the jet near the BH could be measured. (Particles of equally high energy
may exist in $\rho>\rho_1$ but will spread out away from the axis.)

Let us make a brief qualitative remark about the observability of the two
species, (a) and (b), of geodesics outgoing from the ergosphere, studied in
section 3 (after (45)). As illustrated by the figure 2, for each fixed value
of $\rho _{1}$\ there is one (b)-geodesic only, which is the limit of many
(one infinity of) (a)-geodesics when the turning point tends to the infinity
($z_{2}\rightarrow \infty $, $\rho _{2}\rightarrow \rho _{1}$). However, the
(a)-type geodesics, though much more numerous than the (b)-type geodesics,
are, directly or indirectly (i.e. by radiation, if charged), much more
difficult to observe.

Indeed, contrary to the set of (b)-particles framing the jet in one
direction (collimation along the poles), the (a)-particles ejected from the
ergosphere along unbound geodesics at lower latitudes are dispersed into the
whole 3D-space ($4\pi $ steradians).\ The (a)-particles never produce a beam
into one privileged direction but instead dilute in the whole space.
Observed from the infinity in one line of sight ($\theta =$ constant, $\phi =
$constant), one single (a)-particle could directly be detected.While, from
the infinity in the line of sight $z$\ ($\theta =0$, $\forall \phi $), the
observer will see one infinity (each point of the perimeter of the circle of
radius $\rho _{1}$) of (b)-particles. The result is reinforced when we
extend it to all the possible values of $\rho _{1}$. Encircling the foot of
the (b)-jet, the (a)-particles frame a gerb, from the basis of which a
possible indirect effect of isotropic radiation emission (from accelerated
charged particles) could be observed, during the jet eruption.

Besides, by their dispersion, the pressure the (a)-particles locally exert
on the ambient medium is much weaker than the pressure exerted by the
numerous coherent (b)-particles of the jet (a narrow parallel beam is more
incisive). The (a)-particles are probably more rapidly thermalised than the
(b)-particles of the jet. So, one might expect that many particles ejected
at lower latitudes never attain infinity (neither the height of the jet),
and most of them feed the medium, framing a halo around the BH, falling
inside again, or returning to the accretion disk.

We noted also that geodesics with $\mathcal{Q}<0$ can be asymptotic to lines
with constant $\theta$. These asymptotes allow us to define another type of
jet which is bigger and less collimated than the previous one. It is
interesting to remark that recent observations \cite{SheFraWhi03,SauTsiTru02}
suggest the existence of two different types of jets precisely of these
sorts, i.e.\ narrowly and broadly collimated.

There exists an ensemble of geodesics that tend asymptotically to these
conical characteristics. The unbounded geodesics have mainly been discussed,
however, by using Boyer-Lindquist coordinates $r$ and $\theta$ by the
majority of authors. If we rewrite our results, using these coordinates, we
may interpret our results and compare to those of other authors. However, as
we show below, these coordinates are not as well-suited to the issues we
have discussed.

Geodesics with $L_{z}\neq 0$ may reach low values of $\rho /z$, if $b_{2}$
is large enough, but must be bounded away from $|\cos \theta |=1$ (i.e.\ $%
\theta =0$ or $\theta =\pi $), since those values would imply $\dot{\theta}%
^{2}<0$, from (\ref{2}), cf.\ \cite{Cha83}, p.\ 348. In practice this means
that a narrow jet along the axis must be composed of particles with very
small $L_{z}$. Particles with non-zero $L_z$ could only lie within a jet
with bounded $\rho$ for a limited distance, because large enough $z$ would
imply $\dot{\theta}^2<0$. If $\mathcal{Q} \geq 0$, the orbits reverse the
sign of $\dot{\theta}$ and reach the equatorial plane, and would thus be
expected to be absorbed by the accretion disk. For $\mathcal{Q} < 0$ they
are confined to a band of values of $\theta$ given by the roots of $S^2=0$.
These are the `vortical' trajectories of de Felice et al.\ \cite%
{deCal72,deCur92,deCar97}. Depending on the maximum opening angle $\theta$,
these may still hit, and presumably be absorbed by, a thick accretion disk
\cite{deCur92}. Such orbits can be adequately populated by Penrose-like
processes \cite{Wil95,Wil04}, and might undergo processes which reduce the
opening angle \cite{deCur92,deCar97}. A jet composed of such particles would
tend to be hollow and would have a larger radius $\rho$ at large $z$ than is
obtained for orbits with $L_z=0$, and hence be observationally
distinguishable. The presence of these escaping trajectories spiralling
round the polar axis can be associated with the gravitomagnetic effects due
to the rotation of the hole, one of whose consequences is that even curves
with $L_z=0$ have a non-zero $\mathrm{d}\phi/\mathrm{d}t$ at finite
distances.

Thus although an infinitely extended jet of bounded $\rho $ radius would
only contain particles with $L_{z}=0$, which we would expect to be a set of
measure zero among all particles ejected, we shall consider this as a good
model even for real jets. In practice, interactions with other forces and
objects, which would affect the jet both by gravitational and other forces,
have to be taken into account once the jet is well away from the BH, and
these influences might or might not improve the collimation. In \cite%
{deCar97}, the authors discussed possible improved collimation for particles
of low $L_{z}$ using forces which have a timescale long compared with the
dynamical timescale of the geodesics, and which act to move particles to new
geodesics with changed parameters. It should be noted that if the object
producing the jet is modelled as a rotating black hole, production of a
collimated jet only arises naturally if the object throws out energetic
particles with low $L_{z}$, since our discussion shows that other particles
cannot join such a jet unless there is some other strong collimating
influence away from the BH.

However previous authors have not pointed out the existence of asymptotes $%
\rho=\rho_1$, presumably because they are less obvious when using
coordinates $r$ and $\theta$. In fact, considering $z\rightarrow\infty$, the
expressions (\ref{10}) and (\ref{12}) produce $\cos\theta\approx
1-(\rho^2_1/2z^2)+O(z^{-3})$, $\sin\theta\approx (\rho_1/z)+O(z^{-4})$ and $%
r\approx z+1+(\rho^2_1/2z^2)+O(z^{-3})$. With these expansions it is clear
that in the limit $\theta =0$ one would have to take the limit of $%
r\sin\theta$ to allow $\rho_1$ to be determined.

In the same vein, to find the values of asymptotes $z=z_1\neq 0$ near the
equatorial plane $\theta=\pi/2$ for the incoming particles (see (\ref{29}))
one has to study $r\cos\theta$ if one uses the coordinates $r$ and $\theta$.
In fact, one finds for the asymptotic expansion $\rho\rightarrow\infty$ the
following expressions, $\cos\theta\approx (z_1/\rho)+O(\rho^{-3})$, $%
\sin\theta\approx 1-(z_1^2/2\rho^2)+O(\rho^{-3})$ and $r\approx\rho+1+{%
[z_1^2+1-(a/M)^2]/2\rho}+O(\rho^{-3})$.

\section{Conclusion}

Our main results are the following.

There are projections of geodesics all over the meridional planes. Among
these geodesics there are some, with vertical asymptotes parallel to $z$
which can form a perfectly collimated jet. There are, as well, geodesics
with horizontal asymptotes parallel to the radial coordinate $\rho$, that
can represent the paths of incoming particles leaving the accretion disk.

These two types of geodesics have intersection points that can be situated
inside the ergosphere. At these points a Penrose process can take place,
producing the ejection of particles along the axis with bigger energies than
the energies of incoming particles close to the equatorial plane. The
energies of outgoing particles are significantly larger than the ones of the
incident particles for the asymptotically vertical geodesics near the scale $%
a/M$ of the ergosphere diameter in the coordinate $\rho $, so such particles
can show collimation around the surface of a tube of diameter $2a/M$ centred
on the axis of symmetry. Such collimated outgoing particles have to have a
zero orbital momentum $L_{z}=0$, which implies, from the Penrose process,
that the incoming particles have a negative orbital momentum, $L_{z}^{\prime
}<0$. Thus the jet has to be fed from incoming particles with retrograde
orbits in the accretion disk. There is evidence for the existence of
substantial counterrotating parts of accretion disks \cite%
{KoiMeiShi00,ThaRydJor97}, and such counterrotations could explain the
viscosity inducing the instabilities which trigger the falling of matter
towards the ergosphere. It is now known \cite{MirRod94,MirRod99,Mir06} that
there is a close connection between instabilities in the accretion disk and
the genesis of jets for quasars and microquasars.

The most energetic incoming particles are those near the equatorial plane.
Hence the incoming particles which produce the most energetic outgoing
particles by a Penrose process in the ergosphere, whose maximum size is $z_e=%
\sqrt{A}$, are those with angular momentum $L^{\prime}_z<0$ and a very small
impact parameter $z_1$.

Also, the limiting diameter of the core of a perfectly collimated jet
depends upon the size of the ergosphere. The effective thickness of this
part of the jet in this case is of the order of $2\rho_e= 2a/M$.

Our idealised model is based on the well-behaved vacuum stationary exact
solution of Einstein's equations with axial symmetry, namely the Kerr
metrics, which does not take into account the ambient medium. Though this
medium is very dilute, it plays a non-negligible role on the more complex
global scenario for jets like progressive widening of the beam, advent of
knots, lobes, etc... However, for the scenario that we are here concerned,
namely the beginning of the jet (parsec scale for microquasars,while some
hundred parsecs for AGN, depending on the BH mass), where it is strongly
collimated, our approximation of test-particles along geodesics is relevant.
Indeed, the observed jets stemming from active galactic nuclei ejected along
the polar axis have ultrarelativistic speeds, typically $v_{j}=0.99995$c.
The ejected particles, forming the jets, are thermalized with temperatures
of the order $10^{5}$K (\cite{Fill09}) producing a lateral force from the
pressure gradient between the thermal energy of the particles in the outflow
and the low density enveloping medium (\cite{Pun99a} ,\cite{Pun99b}). The
internal particle trajectories to these jets expand laterally at the speed
of sound, being of the order $v_{s}=30$km/s \cite{Fill09}, asymptotically
forming a conical shape with an opening angle of the order of the inverse
Mach number $(v_{s}/v_{j})=10^{-4}$ radians. As we can see (\cite{Pun99b},
Appendix), the more realistic trajectories corresponding to such corrective
terms represent only a small perturbation to the geodesics.

The model that we present to explain the formation and collimation of jets
arises essentially from relativistic strong gravitational field phenomena
without resort to electromagnetic phenomena. From this point of view the
model could be interesting also for understanding observational evidence of
neutral particles emitted from the inner jet itself. For example, the recent
observations of Ultra High Energy Cosmic Rays (difficult to explain,
implying neutral particles such as neutrinos, or H or Fe atoms, etc. \cite%
{Aug07,Aug07a,DerRazFin08}, and \cite{HES09} and references therein) is a
new challenge. To explain the Very High Energy of such neutral (massive)
particles, especially neutrinos \cite{Aug07,Aug07a} (which are able to
travel freely over large distances), our model very naturally suggests that
they could be directly coming from the collimated inner jet, which would
privilege sources (BH) with rotational $z$-axis along the line of sight of
the observation. Massless particles (photons) would be emitted by charged
particles accelerated along the collimated inner jet \cite{DerRazFin08,HES09}%
, which would privilege sources (BH) with rotational $z$-axis perpendicular
to the line of sight of the observation.

Our model is sufficiently general to fit various types of observed jets,
like GRB, jets ejected from AGN or from microquasars, whenever they are
energetic enough to be explained by just a rotating black hole fed by an
accretion disk in an axisymmetric configuration. The main drawback is the
need to preferentially populate the geodesics which can form such collimated
jets. Work is in progress on this question to determine a possible
confrontation of the model with observations.\ Our preliminary studies led
us to understand the fundamental role of the function $P(r)$ of the
geodesics equations (See (2) and (19)). As an example, in the special case
where the equation $P(r)=0$ has a real double root, there exist only two
narrow ranges of $\rho _{1}$ values for large values of $E$. In this case,
we can evaluate from the power, for example of radio loud extragalactic jets
(\cite{Willott99}), or of microquasars jets (\cite{Fend04}), the particle
density, the mean kinetic energy by particle, the mean velocity and the
Lorentz factor of the jets. These results, since they require a long
presentation, deserve a separate paper which is under preparation.

The existence of vacuum solutions of the Einstein equations of Kerr type but
with a richer, not connected, topological configuration of the ergosphere
(see \cite{GarMarSan02}, figs.\ 7 to 10), allows us to propose the existence
of double jets, because they are expected to come out from the ergosphere.
These bipolar jets have been observed (see for instance \cite{SkiMeiBar97},
\cite{SahTraWat98} fig.\ 1, \cite{Far03} fig.\ 2, and \cite{KwoSuHri98}) and
could be naturally interpreted in a generalization of our model.

\textbf{Acknowledgement} We are grateful to Dr.\ Reva Kay Williams for
correspondence concerning her papers and for further references, and to
Prof.\ J. Bi\u{c}\'ak for bringing \cite{BicSemHad93} to our attention.




\begin{thebibliography}{(Bi\u{c}\'{a}k et al., 93)}
\bibitem[(Auger, 2007a)]{Aug07} Auger collaboration 2007, Science,318,939

\bibitem[(Auger, 2007b)]{Aug07a} Auger collaboration 2007, Astropart.
Phys.,29,188, erratum: ibid 30, 45 (2008), arXiv:0712.2843v2.

\bibitem[(Bi\u{c}\'{a}k et al., 93)]{BicSemHad93} Bi\u{c}\'{a}k, J., Semer%
\'{a}k, O. \& Hadrava, P. 1993, Mon. Not. R. Astron. Soc., 263, 545

\bibitem[(Blandford \& Znajek, 77)]{BlaZna77} Blandford, R.D. \& Znajek,
R.L. 1977, Mon. Not. R. Astron. Soc., 179, 433

\bibitem[(Carter, 1968)]{Car68} Carter, B. 1968, Phys. Rev., 174, 1559

\bibitem[(Chandrasekhar, 1983)]{Cha83} Chandrasekhar, S. 1983, The
Mathematical Theory of Black Holes (Oxford: Oxford University Press), 346

\bibitem[(de Felice \& Calvani, 72)]{deCal72} de Felice, F. \& Calvani, M.
1972, Nuovo Cimento B, 10, 447

\bibitem[(de Felice \& Carlotto, 97)]{deCar97} de Felice, F. \& Carlotto,
L., 1997, Astrophys. J., 481, 116

\bibitem[(de Felice \& Curir, 92)]{deCur92} de Felice, F. \& Curir, A. 1992,
Class. Quantum Grav., 9, 1303

\bibitem[(de Felice \& Zanotti, 00)]{deZan00} de Felice, F. \& Zanotti, O.
2000, Gen. Rel. Grav., 32, 1449

\bibitem[(Dermer et al., 08)]{DerRazFin08} Dermer, C.D., Razzaque, S.,
Finke, J.D. \& Atoyan, A. 2008, arXiv:0811.1160v3

\bibitem[(Fargion, 03)]{Far03} Fargion, D. 2003, Puzzling afterglow's
oscillations in GRBs and SGRs: tails of precessing jets, Tech.\ Rep.
astro-ph/0307314, contribution to the Vulcano conference.

\bibitem[(Fender et al., 04)]{Fend04} Fender, R.P.,Belloni, T.M. and Gallo,
E. 2004, MNRAS, 355, 1105.

\bibitem[(Filloux, 09)]{Fill09} Filloux, C. 2009, Ph.D. thesis, Universit%
\'{e} de Nice Sophia-Antipolis, France.

\bibitem[(Gariel et al., 02)]{GarMarSan02} Gariel, J., Marcilhacy, G., \&
Santos, N.O. 2002, Class. Quantum Grav., 19, 2157

\bibitem[(Herrera \& Santos, 07)]{HerSan07} Herrera, L. \& Santos, N.O.
2007, Astrophys. Space Sci., 310, 251

\bibitem[(HESS collab., 09)]{HES09} HESS collaboration 2009,
arXiv:0903.1582v1, to appear in Astrophys.\ J.\ Letters

\bibitem[(Hughson et al., 72)]{HugPenSom72} Hughson, L.P., Penrose, R.,
Sommers, P. \& Walker, M. 1972, Commun.math. phys., 27, 303

\bibitem[(Koide et al., 00)]{KoiMeiShi00} Koide, S., Meier, D.L., Shibata,
K. \& Kudoh, T. 2000, Astrophys. J., 536, 668

\bibitem[(Kwok et al., 98)]{KwoSuHri98} Kwok, S., Su, K.Y.L. \& Hrivnak,
B.J. 1998, Astrophys. J., 501, L117

\bibitem[(Livio, 99)]{Liv99} Livio, M. 1999, Phys. Rep., 311, 225

\bibitem[(McClintock et al., 06 )]{McCShaNar06} McClintock, J.E., Shafee,
R., Narayan, R., Remillard, A., et al. 2006, Astrophys. J., 652, 518 ,
astro-ph/0606076

\bibitem[(Mirabel, 06)]{Mir06} Mirabel, I.F. 2006, Black holes: from stars
to galaxies, Tech.\ Rep. astro-ph/0612188 , concluding Remarks of IAU
Symposium 238: "Black Holes: From Stars to Galaxies -Across the Range of
Masses". {P}rague 14-18 August 2006

\bibitem[(Mirabel \& Rodriguez, 94)]{MirRod94} Mirabel, I.F. \& Rodriguez,
L.F. 1994, Nature, 371, 46

\bibitem[(Mirabel \& Rodriguez, 99)]{MirRod99} Mirabel, I.F.\ \& Rodriguez,
L.F. 1999, Ann. Rev. Astr. Astrophys., 37, 409

\bibitem[(O'Neill, 95)]{ONe95} O'Neill, B.\ 1995, The Geometry of {Kerr}
Black Holes ( Wellesley, Massachusetts: A K Peters Ltd.)

\bibitem[(Opher et al., 96)]{OphSanWan96} Opher, R., Santos, N.O. \& Wang,
A. 1996, J. Math. Phys., 37, 1982

\bibitem[(Penrose, 69)]{Pen69} Penrose, R. 1969,\ Rivista del Nuovo Cimento,
Numero Special 1, 252

\bibitem[(Piran et al., 01)]{PirKumPan01} Piran, T., Kumar, P., Panaitescu,
A. \& Piro, L. 2001, Astrophys. J., 560, L167

\bibitem[(Piran \& Shaham, 77)]{PirSha77} Piran, T. \& Shaham, J. 1977,
Phys. Rev. D, 16, 1615

\bibitem[(Punsly, 99a)]{Pun99a} Punsly, B. 1999a, Astrophys. J., 527, 609.

\bibitem[(Punsly, 99b)]{Pun99b} Punsly, B. 1999b, Astrophys. J., 527, 624.

\bibitem[(Punsly, 01)]{Pun01} Punsly, B. 2001, Black Hole
Gravitohydromagnetics (Berlin and Heidelberg : Springer-Verlag)

\bibitem[(Punsly \& Coroniti, 90a)]{PunCor90a} Punsly, B. \& Coroniti, F.V.
1990a, Astrophys. J., 354, 583

\bibitem[(Punsly \& Coroniti, 90b)]{PunCor90} Punsly, B. \& Coroniti, F.V.
1990b, Astrophys. J., 350, 518

\bibitem[(Rees et al., 76)]{ReeRufWhe76} Rees, M., Ruffini, R. \& Wheeler,
J.A. {1976,} Black Holes, Gravitational Waves and Cosmology: An Introduction
to Current Research (New york: Gordon and Breach Science Publishers)

\bibitem[(Sahai et al., 98)]{SahTraWat98} Sahai, R., Trauger, J.T., Watson,
A.M., et~al. 1998, Astrophys. J., 493, 301

\bibitem[(Sauty et al., 02)]{SauTsiTru02} Sauty, C., Tsinganos, K.
\&Trussoni, E. 2002, in Springer Lecture Notes in Physics, Vol. 589 ,
Relativistic Flows in Astrophysics, edited by A.W.Guthmann,
M.Georganopoulos, A.Marcowith \& K.Manolakou (Berlin and Heidelbergn:
Springer-Verlag), 41, astro-ph/0108509

\bibitem[(Sharp, 79)]{Sha79} Sharp,\ N.A.\ 1979,\ Gen. Rel. Grav., 10, 659

\bibitem[(Sheth et al., 03)]{SheFraWhi03} Sheth, K., Frail,\ D.A., White,
S., et al. 2003, Astrophys. J., 595, L33

\bibitem[(Skinner et al., 97)]{SkiMeiBar97} Skinner, C.J., Meixner, M.,
Barlow, M.J., et al., 1997, Astron. and Astrophys., 328, 290

\bibitem[(Stewart \& Walker, 74)]{SteWal74} Stewart, J.M. \& Walker, M.
1974, Springer Tracts in Modern Physics, Vol.69,\ Black holes: the outside
story ( Berlin: Springer)

\bibitem[(Thakar et al., 1997)]{ThaRydJor97} Thakar, A.R., Ryden, B.S.,
Jore, K.P. \& Broeils, A.H. 1997,\ Astrophys. J., 479, 702

\bibitem[(Willott et al., 99)]{Willott99} Willott, C.,Rawlings, S.,Blundell,
K. and Lacy, M. 1999, MNRAS, 309, 1017.

\bibitem[(Wilkins, 72)]{Wil72} Wilkins, D.C. 1972, Phys. Rev. D, 5, 814

\bibitem[(Williams, 95)]{Wil95} Williams, R.K. 1995, Phys. Rev. D, 51, 5387

\bibitem[(Williams, 04)]{Wil04} Williams, R.K. 2004, Astrophys. J., 611, 952
\end{thebibliography}
\end{document}